\documentclass[twocolumn]{aastex631}
\usepackage{amsmath}
\usepackage{threeparttable} 

\usepackage{bm}
\usepackage{xcolor}

\begin{document}

\title{Pitch-Angle Scattering of Cosmic Rays: {Confronting Theory with Observations}}

\author[0000-0003-2560-8066]{Huirong Yan}

\affiliation{Deutsches Elektronen Synchrotron (DESY), Platanenallee 6, D-15738 Zeuthen, Germany}
\affiliation{Institut für Physik und Astronomie, Universität Potsdam, D-14476 Potsdam, Germany}

\author[0000-0003-4268-7763]{Siqi Zhao}
\affiliation{Institut für Physik und Astronomie, Universität Potsdam, D-14476 Potsdam, Germany}
\affiliation{Institute of Science and Technology for Deep Space Exploration, Nanjing University, China}

\author[0000-0003-3529-8743]{Ming Zhang}
\affiliation{Florida Institute of Technology, 150 W University Blvd, Melbourne, FL 32901, USA}

\correspondingauthor{Huirong Yan}
\email{huirong.yan@desy.de}

\begin{abstract}
Cosmic ray (CR) propagation is controlled by scattering in turbulent magnetic fields in space. In general, diffusive propagation is governed by pitch-angle diffusion in phase space. In this study, pitch-angle diffusion in the local interstellar medium (LISM) deduced from the analysis of {the CR small-scale anisotropy data} from the Tibet AS$\gamma$ experiment is compared with theoretical predictions. While it is difficult to reconcile the inferred LISM pitch angle diffusion coefficient with conventional theoretical results of particle scattering by Alfv\'{e}nic turbulence, we find remarkable agreement with the predictions of particle scattering by quasi-slab fast modes whose properties are shaped by damping in the warm ionized medium. These findings offer direct evidence that CR scattering is predominantly governed by fast-mode turbulence. Furthermore, the comparison between experimental and theoretical results imposes strong constraints on plasma and magnetic field parameters within the local bubble, indicating that the LISM is in a low $\beta$ condition. The turbulence in the LISM should be compressible with an amplitude in the range of {\bf $
0.3\lesssim  \delta B/B_0 <1$}. 
\end{abstract}


\section{Introduction}
It is challenging to directly trace the behavior of cosmic ray (CR) pitch-angle scattering in the interstellar medium, although numerous studies using theoretical calculations and test-particle simulations have been conducted. The difficulty arises because it requires determining the pitch-angle distribution of CRs in the pristine local interstellar medium (LISM), whose anisotropy is extremely small, on the order of $\lesssim 10^{-3}$ \citep[e.g.,][]{Longair1997, 2010ApJ...718L.194A, 2010ApJ...711..119A, 2017ApJ...842...54L, 2019ApJ...871...96A, ZhangM2020,MZ2025, Qiao2026}. While the source distribution influences large-scale anisotropy \cite[e.g.,][]{Blasi2012, Liu2017}, the small-scale features are primarily determined by the pitch-angle scattering rate $D_{\mu\mu}$ \citep{MZ2025}. $D_{\mu\mu}$ is governed by magnetic fluctuations instead of the source distribution, as demonstrated in the literature \cite[e.g.][]{Giacinti2012, Ahlers2017}. The CR anisotropy is also subject to strong distortion by the heliosphere and must be eliminated in order to reconstruct the particle distribution in the LISM.

Over the past decade, our understanding of the heliosphere has improved significantly through observations by Voyager and the Interstellar Boundary Explorer (IBEX). These measurements have enabled increasingly sophisticated magnetohydrodynamic (MHD) modeling of the heliospheric structure, making it possible to map particle distributions observed at Earth back to the pristine LISM. In particular, recent work has demonstrated that the pitch-angle distribution {in the LISM} can be reconstructed by backtracking particle trajectories through the heliospheric plasma and magnetic field \citep{ZhangM2020, MZ2025}.

The pitch-angle diffusion coefficient can be derived from the LISM distribution {\bf under the diffusion approximation} \citep[see][and references therein]{MZ2025}. The resulting observationally inferred pitch-angle diffusion coefficients provide the foundation for the present study and enable, for the first time, a direct comparison between cosmic-ray observations and theoretical predictions. In this paper, we quantitatively compare these inferred diffusion coefficients with the predictions of the nonlinear theory (NLT) model for compressible turbulence {\bf in the diffusion regime} developed by \cite[][hereafter YL08]{Yan2008}, using the current understanding of interstellar turbulence as the theoretical framework.

\section{Pitch-angle scattering from theory and observations}

The pitch-angle scattering of CRs is governed by their interactions with MHD fluctuations \citep[see][]{Jokipii1966, Schlickeiser2002, Yan2022}. Depending on the energy of CRs, two main sources of MHD-scale perturbations contribute. For average CR flux in the interstellar medium, CRs with energies $\lesssim 100$ GeV are predominantly scattered by self-generated MHD waves arising from streaming and pressure-anisotropy instabilities \citep{FG04, Yan2004, Yan2011,Bykov2013,Lebiga2018}. For CRs with energies $\gtrsim 100$ GeV, scattering by preexisting turbulence is dominant. The physical reasons for the energy dependence are two-fold. First, the small-scale instabilities operate only when there is sufficient particle flux for the wave growth to overcome various damping processes, including damping by background turbulence \citep{FG04,Yan2004,Yan2011,Lazarian2016} and nonlinear Landau damping \citep{Lee1973, Kulsrudbook}. Second, preexisting large-scale turbulence -- particularly fast modes, which are most effective for scattering those $\gtrsim 100$GeV CRs—are subject to damping and become subdominant for CRs with energies $\lesssim 100$ GeV (YL08). 

\begin{figure*}[ht]
\centering
\gridline{
\fig{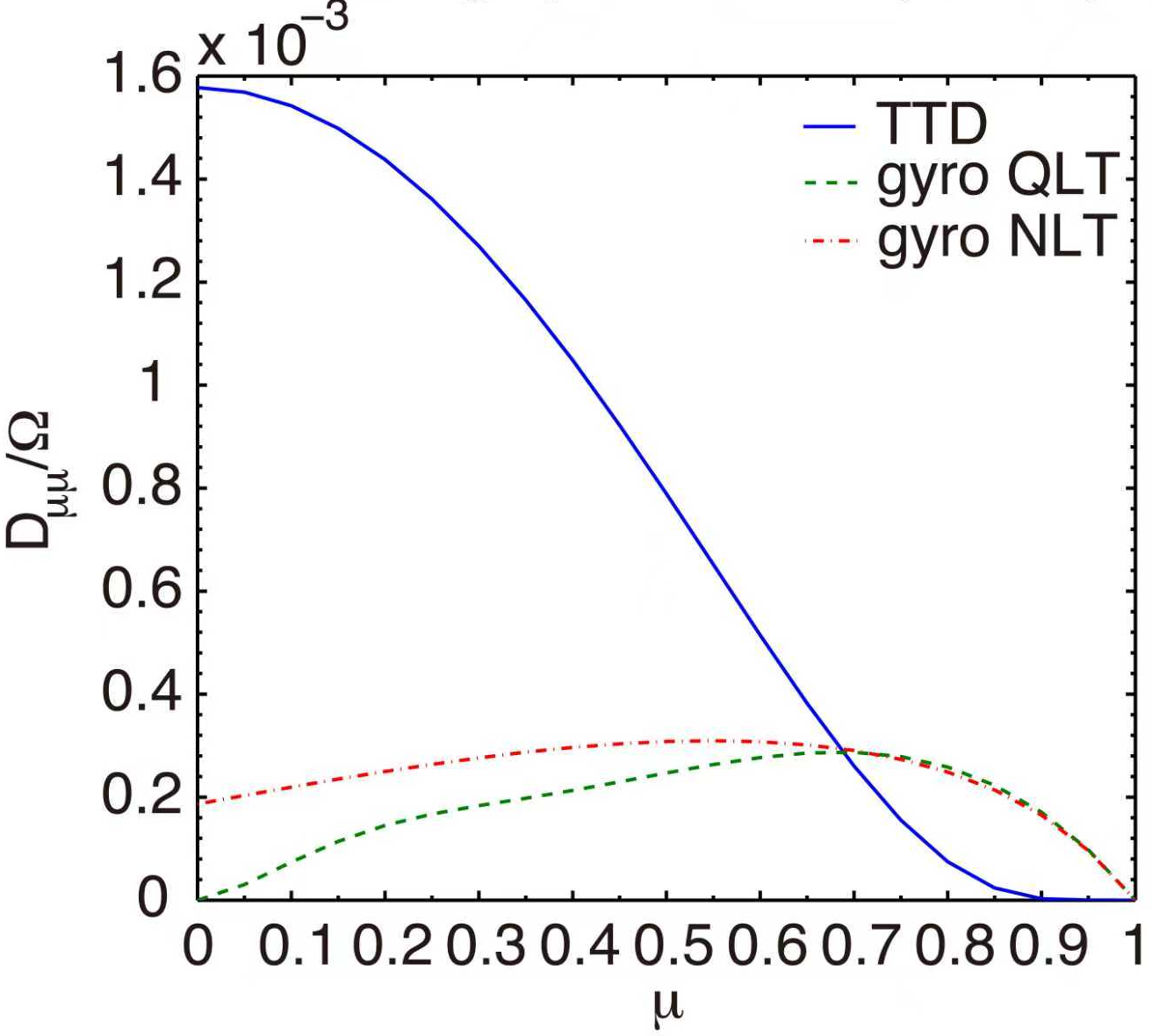}{0.45\textwidth}{(a)}
\fig{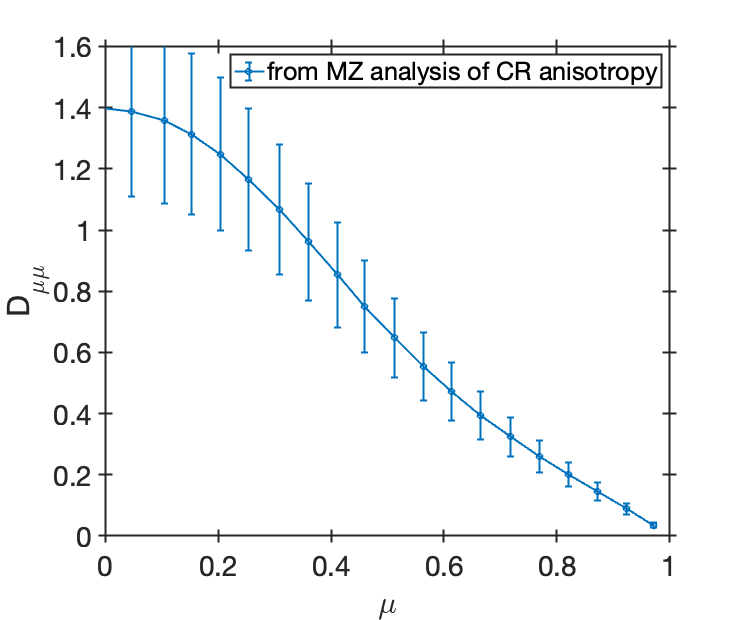}{0.48\textwidth}{(b)}}
    \caption{Comparison of cosmic ray scattering rates in warm ionized medium predicted by YL08 and values obtained from CR anisotropy analysis from \citep{MZ2025}. {\bf The error bars were estimated by allowing the overall slope of the pitch-angle distribution of the scattering rate $\nu^s$ to vary by approximately 20\%, while still consistent with the observed scatter of the data points.}}
    \label{fig:originalones}
\end{figure*}

For the TeV cosmic rays (CRs) examined in this study, scattering is governed exclusively by large-scale magnetohydrodynamic (MHD) turbulence. A central result derived from the analysis of TeV CR anisotropy measurements is the weak dependence of the scattering rate $\nu^{\mathrm{s}}$ on the particle pitch-angle cosine $\mu$ \citep{MZ2025}, which is expressed as:
\begin{equation}
    \nu^s = D_{\mu\mu}^s/(1-\mu^2)=(D_0 (224  \cos(2 \pi \mu / 1.73) + 564) / 564),
    \label{exp_nuval}
\end{equation}
{\bf where $D_0$ is an adjustable normalization factor with unit $s^{-1}$}. The result is in conflict with the scattering in isotropic or Alfv\'enic turbulence. Instead, the pitch-angle scattering rate obtained from \cite{MZ2025} appears to be consistent with earlier theoretical results obtained for the Warm Ionized Medium (WIM) (YL08) with a temperature close to that of the LISM (see Fig.~\ref{fig:originalones}). This striking agreement motivates this detailed investigation of pitch-angle scattering that explicitly incorporates recent advances in compressible MHD turbulence studies and charged-particle transport.

Compressible modes have been demonstrated to play a pivotal role in particle scattering \citep{Yan2002,Yan2004,Yan2008,Beresnyak2011,Maiti2022}. Alfv\'enic turbulence, which has traditionally been regarded as the dominant contribution, is highly anisotropic \citep{GS95, Chandran2000, Yan2002}, leading to ineffective scattering due to strong suppression of gyroresonance interactions with particles. In contrast, compressible turbulence can provide in addition to the gyroresonance another interaction channel {\bf with cosmic rays} via magnetic mirroring or transit-time damping (TTD) \citep{Schlickeiser1993, Yan2008, Beresnyak2011, YanK2026}. 

\section{Properties of compressible turbulence}

Before revisiting the detailed calculation of CR scattering, we first summarize the relevant properties of compressible turbulence. Compressible turbulence is governed by both cascade and various damping mechanisms. Although substantial progress has been made in understanding compressible turbulence \citep{CL02, CL03, Lazarian2009, Makwana2020}, its fundamental characteristics have long remained a subject of debate. A central issue revolves around the nature of the cascade of fast magnetosonic modes, specifically, whether it actually conforms to the weak Iroshnikov–Kraichnan \cite[IK,][]{Iroshnikov,Kraichnan1965}–type acoustic turbulence model {\footnote{We adopt a balanced turbulence model in the paper, i.e., with zero cross helicity.}.

Recently, significant progress has been made in understanding compressible turbulence across multiple scales through theoretical developments, MHD simulations, and analyses of in situ measurements in the heliosphere \citep[e.g.,][]{Chandran2005,Hadid2018,Zhao2024ApJ}. Theoretical studies indicate that fast modes undergo a weak cascade \citep{Galtier23}. This is supported by the weak nonlinearity revealed by spatiotemporal analysis \citep{Yuen2025,Zhao2026}. Kinetic simulations also reveal an isotropic cascade with a 1D power law spectrum with a slope less than 2 and a scaling of {\bf $SF_2(r)\propto r^{\sigma}$ for the second order structure functions $SF_2$ with $\sigma\sim 0.5$ in both directions parallel and perpendicular to the magnetic field} \citep{Hou2025}. The $k^{-2}$ power law spectrum, reported in some MHD simulations, is not contradictory because of the highly intermittent nature of the steepening. $k$ is defined as the wave number. Furthermore, observations from the solar wind to Earth’s magnetosheath favor a $k^{-3/2}$ spectrum over a $k^{-2}$ spectrum \citep{Zhao2022,Zhao2026}. 

Damping plays a central role in shaping the spectrum and anisotropy of fast-mode turbulence in the inertial range. Early theoretical work suggested that the average propagation angle of fast modes decreases along the cascade, as oblique and quasi-perpendicular modes are preferentially damped through transit-time damping (TTD) {\bf with thermal particles} \citep{YLD2004,Yan2008, Petrosian2006}. The damping rate for TTD at $\beta<1$ can be written as \citep{ginzburg1962, Yan2002} 
\begin{equation}\label{eq_damping}
\begin{aligned}
\Gamma ={}& \frac{\sqrt{\pi\beta}}{4}\frac{\sin^2\theta}{\cos\theta}
\left[\sqrt{\frac{m_e}{m_p}}
\exp\!\left(-\frac{m_e}{m_p\beta\cos^2\theta}\right)\right. \\
&\left. + 5 \exp\!\left(-\frac{1}{\beta\cos^2\theta}\right)\right] k v_f .
\end{aligned}
\end{equation}
where $v_f$ is the fast wave speed. $m_p,m_e$ are the masses of protons and electrons, respectively.  By equating the damping rate with the cascading rate $\nu_{cas}\sim \sqrt{k/L} (V^2/v_f)$, the truncation scale of turbulence can be obtained \citep{Yan2004}. Clearly, the damping is highly dependent on the obliquity of the modes as illustrated in Fig.\ref{fig:dampingcurve}. 
Note that collisionless damping caused by direct interactions between waves and particles always exists, whereas viscous damping only operates on scales larger than the Coulomb mean free path of the medium. 
Direct evidence for this picture was elusive until recently. Particle-in-cell (PIC) simulations provide the first clear evidence of the highly pitch-angle-dependent damping profile \citep{Hou2025,Hou2026}. Moreover, analyses of turbulence in the solar wind and Earth’s magnetosheath demonstrate that compressible fluctuations become increasingly concentrated at larger $k_\parallel$, suggesting that fast modes evolve toward a quasi-parallel configuration at smaller scales \citep{Zhao2024ApJ, Zhao2026_slab}. 

There would be no gyroresonance interaction to bring particles into the normal diffusion regime if the turbulence cascade is truncated on large scales, i.e., $k_c r_L\ll 1$, where $k_c$ is the cutoff wavenumber, and $r_L$ is the particle Larmor radius. 
The quasi-parallel structure of fast modes, which naturally forms on small scales in low $\beta$ plasmas, is essential since the gyroresonance only requires the existence of turbulence perturbations at $k_{\|,res}= \Omega/|v_\|\pm v_A|$, where $\Omega$ is the Larmor frequency, $v_\|, v_A$ is the particle speed along the magnetic field and Alfv\'en speed, respectively. In other words, quasi-2D ($k_\perp\gg k_\|$, $k_\|, k_\perp$ are the projections of the wave vector parallel and perpendicular to the magnetic field) structure contributes marginally to the gyroresonance, whereas quasi-slab modes dominate the interaction \citep[see][]{Yan2002}. Here, plasma $\beta$ is the ratio of the plasma pressure to magnetic pressure.   

In what follows, we will demonstrate that the pitch-angle dependence of the CR scattering rate is directly correlated with the quasi-parallel structure of fast modes.

\begin{figure*}[ht]
\centering
\gridline{
\fig{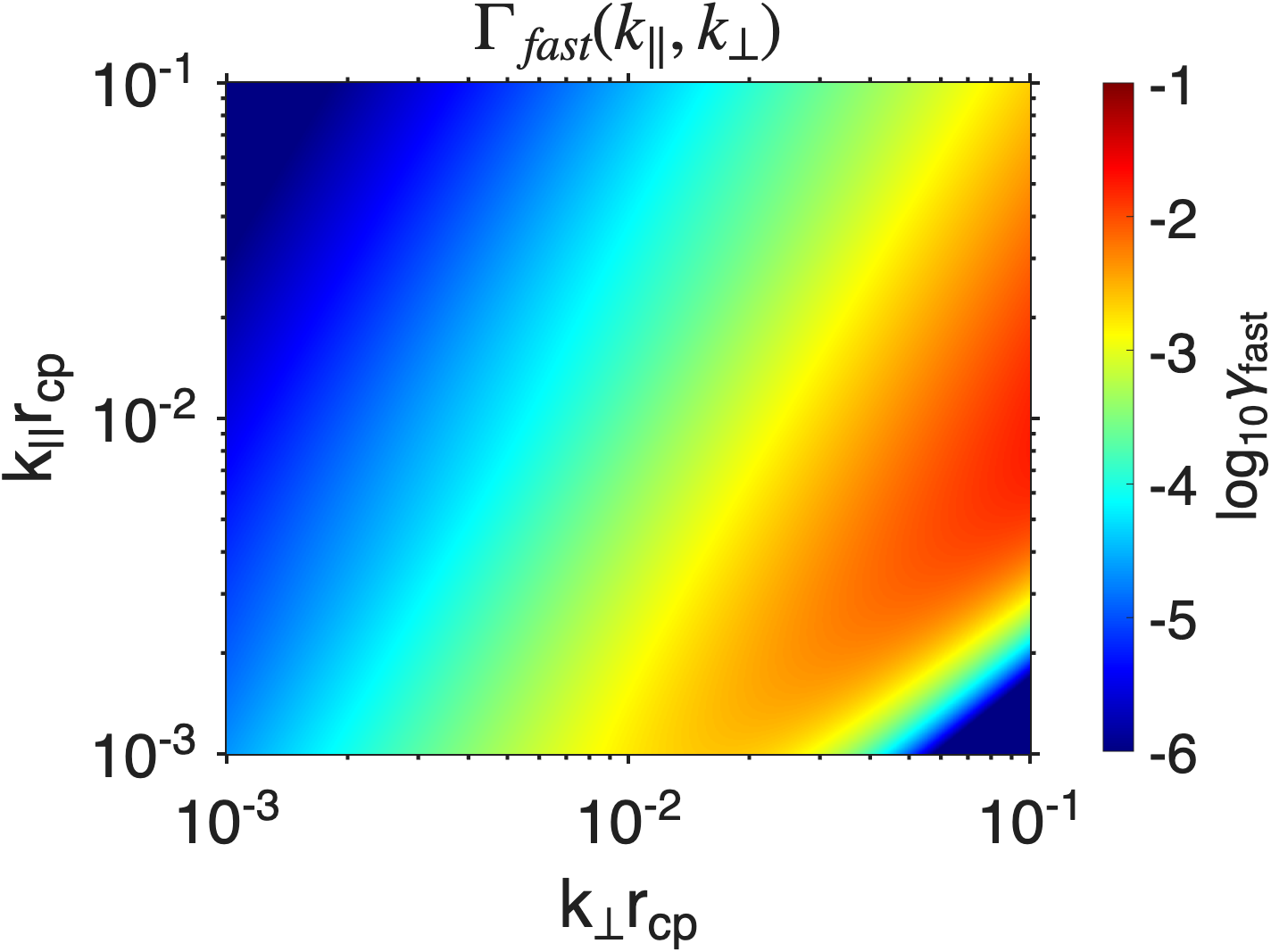}{0.45\textwidth}{(a)}
\fig{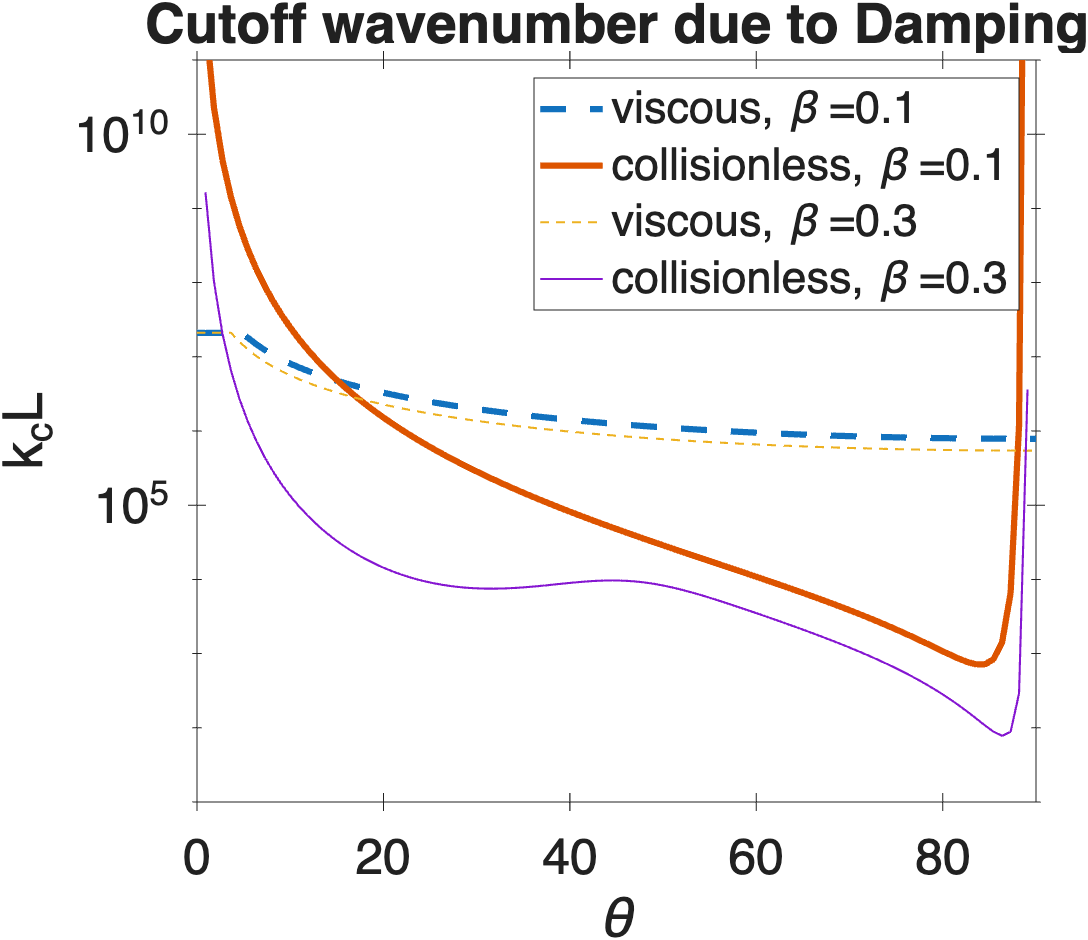}{0.4\textwidth}{(b)}}
    \caption{(a) Collisionless damping rate of fast modes vs. the wavenumbers $k_\perp$ and $k_\|$ normalized by the Larmor radius of protons $r_{cp}$ with $\beta\sim 0.16$; (b) Cutoff scale of fast modes turbulence due to viscous and collisionless damping in a warm ionized plasma as a function of the wave propagation angle to the magnetic field direction.}
    \label{fig:dampingcurve}
\end{figure*}

\section{Pitch-angle scattering by fast-mode turbulence}

Pitch-angle diffusion can be driven by gyroresonance and mirror/TTD interaction. The diffusion coefficient depends on the pitch angle of the particle. For small pitch angles, gyroresonance dominates. For large pitch angles, particularly near $90^\circ$, the mirror interaction is the key scattering mechanism by lifting particles out of the {mirror} potential well \citep{Felice2001,Yan2008}. The TTD treatment is equivalent to that of the mirror interaction in the case of quasi-slab modes with $k_\perp \ll \Omega/v \sim k_{\|, res}$ \citep{YanK2026}. These modes are characterized by a three-dimensional (3D) turbulence spectrum, with power concentrated on $k_\perp\ll k_\parallel$, for which the compression/mirror force is reduced due to the smaller obliquity of the waves. Such quasi-parallel fast modes are expected in low-$\beta$ plasmas ($\beta\lesssim 1$), where the damping rate increases with the wave pitch angle, in general \citep{Yan2004, Yan2008, Petrosian2006, Suzuki2006, Zhao2024ApJ, Hou2025, Zhao2026_slab} (see Fig.~\ref{fig:dampingcurve}). 

\subsection{Magnetic mirror/TTD interaction} 

The diffusion coefficient of the cosine of the particle pitch angle $\mu$ driven by compressible modes can generally be expressed as: \citep[e.g.,][]{Volk1975, Schlickeiser1993, Yan2008}:
\begin{equation} \label{eq1}
\begin{aligned}
D_{\mu\mu}^{\mathrm{TTD}} &=
\frac{\Omega^2 (1 - \mu^2)}{B_0^2}
\int d^3 k \,
R_0(\omega - k_\parallel v\mu)\,
\frac{k_\parallel^2}{k^2} \\
&\left[J_0^{\prime}(W)\right]^2 \,
I^{(M)}(\mathbf{k}),
\end{aligned}
\end{equation}
where $W = k_\perp v_\perp/\Omega$ is the gyro-argument, $J_0^{\prime}$ is the derivative of the 0-th order Bessel function, 
$R_0$ is the 0-th resonance function, and $I^{(M)}(\mathbf{k})$ is the 3D magnetic energy spectrum of compressible modes. In the case of damped fast modes (Fig.~\ref{fig:dampingcurve}) with $k$ or $k_\perp \ll \Omega/v_\perp = r_L$, $W \ll 1$, the Bessel function can be expanded as $ J_0^{\prime}(W) \;\simeq\; \frac{W}{2}$, so that the mirror and TTD can be considered equivalent and
Eq.~(\ref{eq1}) reduces to \citep{Yan2008, YanK2026}:
\begin{equation}
\begin{aligned}
D_{\rm \mu\mu}^{\mathrm{TTD}}
&=
\frac{v_\perp^2 (1 - \mu^2)}{4 B_0^2}
\int d^3 k \,
k_\parallel^2\,
R_0(\omega - k_\parallel v\mu)\\
&\frac{k_\perp^2}{k^2}\,
I^{(M)}(\mathbf{k}),
\label{resfunc}
\end{aligned}
\end{equation}
where the 3D energy spectrum of compressible modes is given by {$I^{(M)}=\delta B^2 (kL)^{-\alpha}$}, and $L$ is the turbulence injection scale.

We adopt the nonlinear formalism in (YL08), which warrants a finite pitch-angle scattering rate throughout the pitch-angle range, including $90^\circ$. The nonlinear resonance function 
\begin{equation}
\begin{aligned}
&R_n(k_{\parallel}v_{\parallel}-\omega\pm n\Omega)\\
&=\Re\int_0^\infty dt e^{i(k_\|v_\|+n\Omega-\omega) t-\frac{1}{2}k_\|^2v_\bot^2t^2 \left(\frac{\langle\delta B_\parallel^2\rangle}{B_0^2}\right)^{1/2}}\\
&=\frac{\sqrt{\pi}}{|k_\|\Delta v_\||}\exp\left[-\frac{(k_\|v \mu-\omega+n\Omega)^2}{k_\|^2\Delta v_\|^2}\right]\\
&\simeq\frac{\sqrt{\pi}}{|k_\||v_\bot M_A^{1/2}}\exp\left[-\frac{(k_\|v \mu-\omega+n\Omega)^2}{k_\|^2v^2(1-\mu^2)M_A}\right],
\label{Rn}
\end{aligned}
\end{equation}
where $M_A\equiv\delta B/B_0$ is the normalized amplitude of the compressible mode perturbation, and $B_0$ is the mean field strength. Inserting it into Eq.~(\ref{resfunc}), we get {\bf for the mirror/TTD interaction}
\begin{equation}
D_{\mu\mu}^{TTD}=D_{\mu\mu}^{\mu=0}\cdot(1-\mu^2)^{3/2} \exp\left[\frac{-\mu^2}{(1-\mu^2)M_A}\right],
\label{eq:TTD_mu}
\end{equation}
where 
\begin{equation}
D_{\mu\mu}^{\mu=0}=\frac{\sqrt{\pi}M_A^{3/2}v}{4L}\int d\xi [k_c(\xi)L]^{-\alpha+4} \xi (1-\xi^2).
\label{eq:TTD_ana}
\end{equation}
The peak of the contribution is from those modes with intermediate obliquity, at $\theta\sim 45^\circ$. Turbulence across all scales {in the inertial range of compressible turbulence} contributes to the mirror interaction, in contrast to gyroresonance. The value crucially depends on the damping profile of the compressible modes, i.e., the cutoff wave number $k_c(\xi)$ (see Fig.~\ref{fig:dampingcurve}b).  Eqs.~(\ref{eq:TTD_mu}, \ref{eq:TTD_ana}) are generally applicable in the presence of damping. Even if the damping is isotropic, which is the case for slow modes or fast modes in partially ionized plasmas, the contribution from mirror/TTD is still finite.

\begin{figure*}[ht]
\centering
\gridline{
\fig{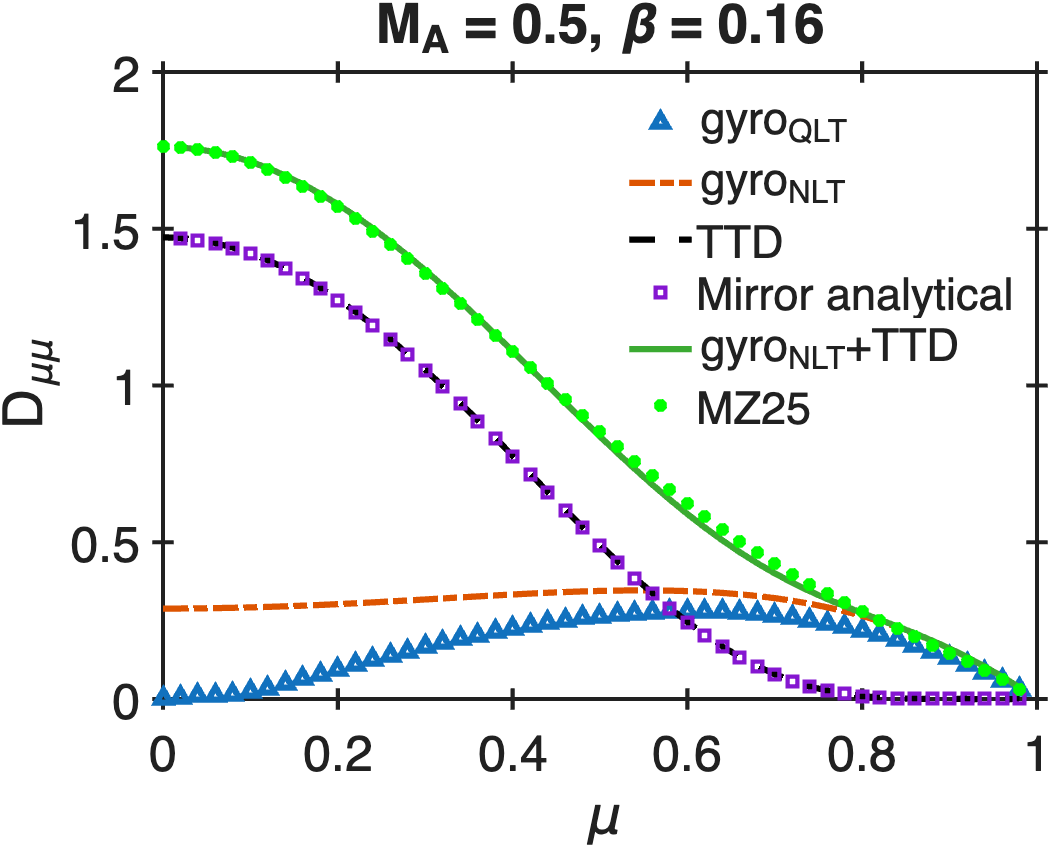}{0.3\textheight}{(a)}
\fig{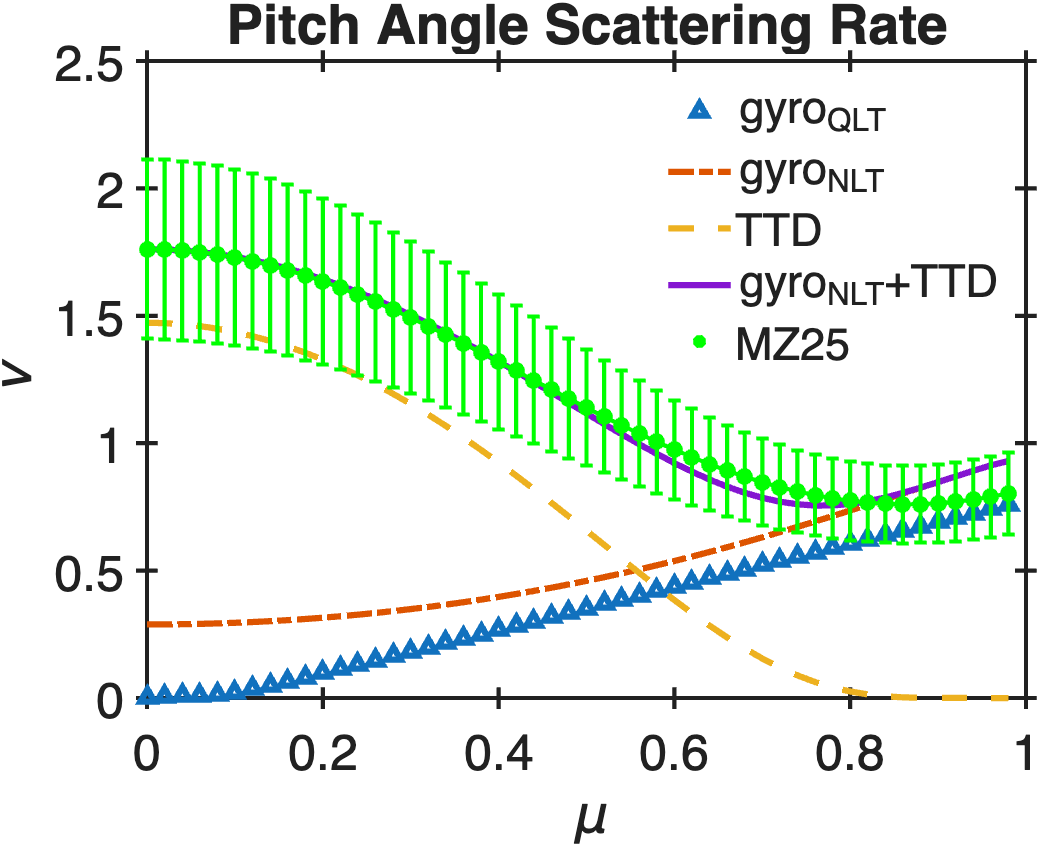}{0.3\textheight}{(b)}}
    \caption{Comparisons of (a) $D_{\mu\mu}$ and (b) scattering rate $\nu= D_{\mu\mu}/(1-\mu^2)$ between theoretical calculations and observation. In both panels, the dash-dot lines represent the contribution from the NLT gyroresonance result, and the triangle line marks the QLT result from gyroresonance. The total $D_{\mu\mu}$ is represented by the solid lines, and the `\textcolor{green}{*}' lines are the experimental results from \citep{MZ2025}. The `\textcolor{violet}{o}' line refers the analytical result for the mirror interaction from Eqs.(\ref{eq:TTD_mu}, \ref{eq:TTD_ana}).}
    \label{fig:comparison}
\end{figure*}

\subsection{Gyroresonance}

Compressible turbulence also efficiently drives particle pitch-angle scattering through gyroresonance. This part of the pitch-angle diffusion coefficient can be given by
\begin{equation}
 \begin{aligned}
D_{\mu\mu}^{\mathrm{G}} &=
\frac{\Omega^2 (1 - \mu^2)}{B_0^2}
\int d^3 k \sum_{n=1}^{\infty} R_n(\omega - k_\parallel v\mu\pm n\Omega)\\
&\frac{k_\parallel^2}{k^2}\left[J_n^{\prime}(W)\right]^2\,
I^{(M)}(\mathbf{k}),
\end{aligned}
\end{equation}
where $J_n^\prime$ is the derivative of the n-th ($n=\pm 1,2,...$) order Bessel function and $R_n$ is the n-th gyroresonance function. For a typical turbulence spectrum as given earlier, the pitch-angle diffusion is dominated by the first gyroresonance. 

The impact of nonlinear broadening on gyroresonance is limited. It is illustrated both in Fig.~\ref{fig:comparison} and in YL08 that the difference between the quasilinear theory (QLT) and the nonlinear theory (NLT) is marginal as the interaction occurs on small scales. The resonance function $R_n(\omega - k_\parallel v\mu\pm n\Omega)$ is a $\delta$ function in the QLT approximation. The expression for gyroresonance in QLT is substantially simplified to the following expression in the case of quasi-slab modes \citep{Yan2004, Yan2008}:
\begin{equation}
\begin{aligned}
    D^{G}_{\mu\mu}&=\frac{M_A^2 \Omega (R\mu)^{\alpha-3}(1-\mu^{2})}{\alpha} \\
&\left[[1+(R\mu)^2]^{-\alpha/2}-(\tan^{2}\theta_c+1)^{-\alpha/2}\right],
\end{aligned}
\label{lbgyro}
\end{equation}
where $\tan\theta_c={k_{\perp,c}}/{k_{\parallel,res}}$, with $k_{\parallel,res} = \Omega/|v\mu \pm v_A|$, $k_{\perp,c}$ is the corresponding cutoff wavenumber for $k_{\parallel}$ fixed at $k_{\parallel,res}$, and $R=v/(\Omega L)$ is the dimensionless rigidity of the CRs.

\subsection{Results of comparison}

Fig.~\ref{fig:comparison} compares theoretical curves with inferred values from the observational analysis of anisotropy data in \cite{MZ2025}, given by Eq.(\ref{exp_nuval}). The behavior of the $\mu$-dependence at $\mu \lesssim 0.8$ can be accurately reproduced by the joint action of mirror/TTD interaction and gyroresonance with fast modes. The agreement provides strong evidence for the dominant role played by quasi-slab fast-mode turbulence on CR scattering and transport. The physical parameters adopted for the LISM are similar to those used in YL08, i.e., $\beta \equiv P_{\text{gas}}/P_{\text{mag}} \simeq 0.16$, $T = 7500$ K, and $n_{\text{gas}} = 0.1$ cm$^{-3}$. These parameters are crucial to determine the damping scales. The overall damping exhibits its strongest dependence on the plasma $\beta$. For a fixed $\beta$ value, variations in other parameters only slightly alter the damping curve as shown in Fig.\ref{fig:dampingcurve}.

For fast-mode turbulence, we set $M_A = 0.5$ in Fig.\ref{fig:comparison}. Note that this is the perturbation in fast modes only. The total turbulence fluctuation should have a larger amplitude. The highly compressible nature of turbulence is consistent with the recent analysis of Voyager 1 measurements \citep{ZhaoL2025}.

The overall $\mu$-dependence places a strong constraint on $M_A$ and the plasma $\beta$ in the LISM. An increase in $M_A$ would increase the contribution of gyroresonance more than that of TTD, resulting in a flatter curve than the one reported in \cite{MZ2025}, as shown in Eq.~(\ref{eq:TTD_ana}). Conversely, increasing plasma $\beta$, for instance, would enhance the damping and reduce the contribution from gyroresonance more dramatically and lead to a steeper decline of the scattering rate with $\mu$. Vice versa. To quantitatively assess the sensitivity of the result on the two key parameters, we conducted the parameter scan by systematically varying both $M_A$ and $\beta$, see contour map in figure \ref{fig:scan_results}. While the analysis clearly indicates that a low-$\beta$ regime is preferred, the admissible interval of $\beta$ is broader, judging from the contour map alone. Nonetheless, we do not believe a high $M_A$ value is favorable since the $M_A$ here only counts for the compressible component of the total fluctuation, which amounts to $\sim 1$ according to independent studies such as \cite{ZhaoL2025}. Assuming $M_A\lesssim 0.5$, beta would be $\lesssim 0.2$. In contrast, the outcomes show only a weak dependence on the turbulence injection scale. Varying the scale from 5 pc to 50 pc yields comparable results.

\begin{figure*}[ht]
\centering
\gridline{
\fig{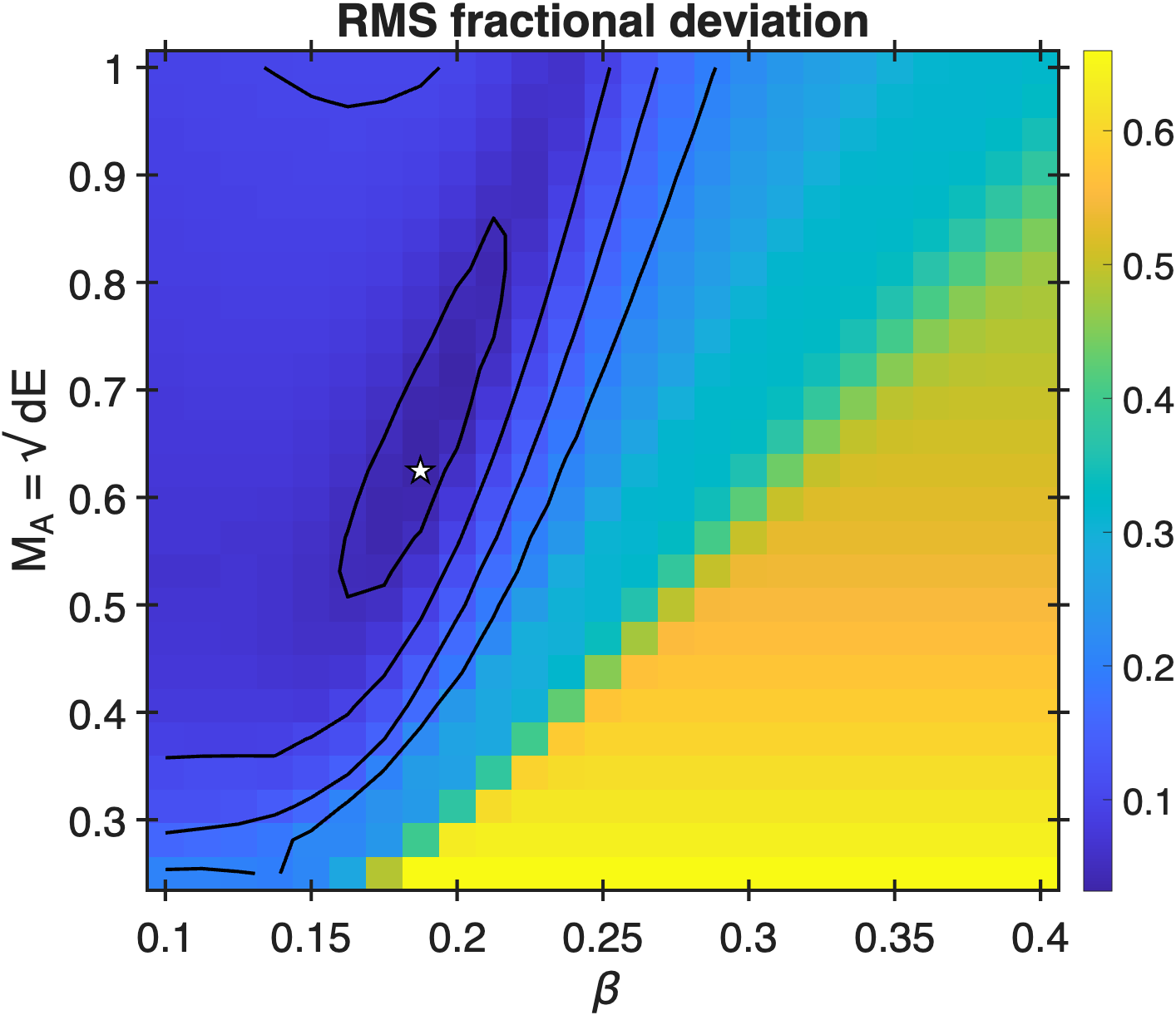}{0.3\textheight}{(a)}
\fig{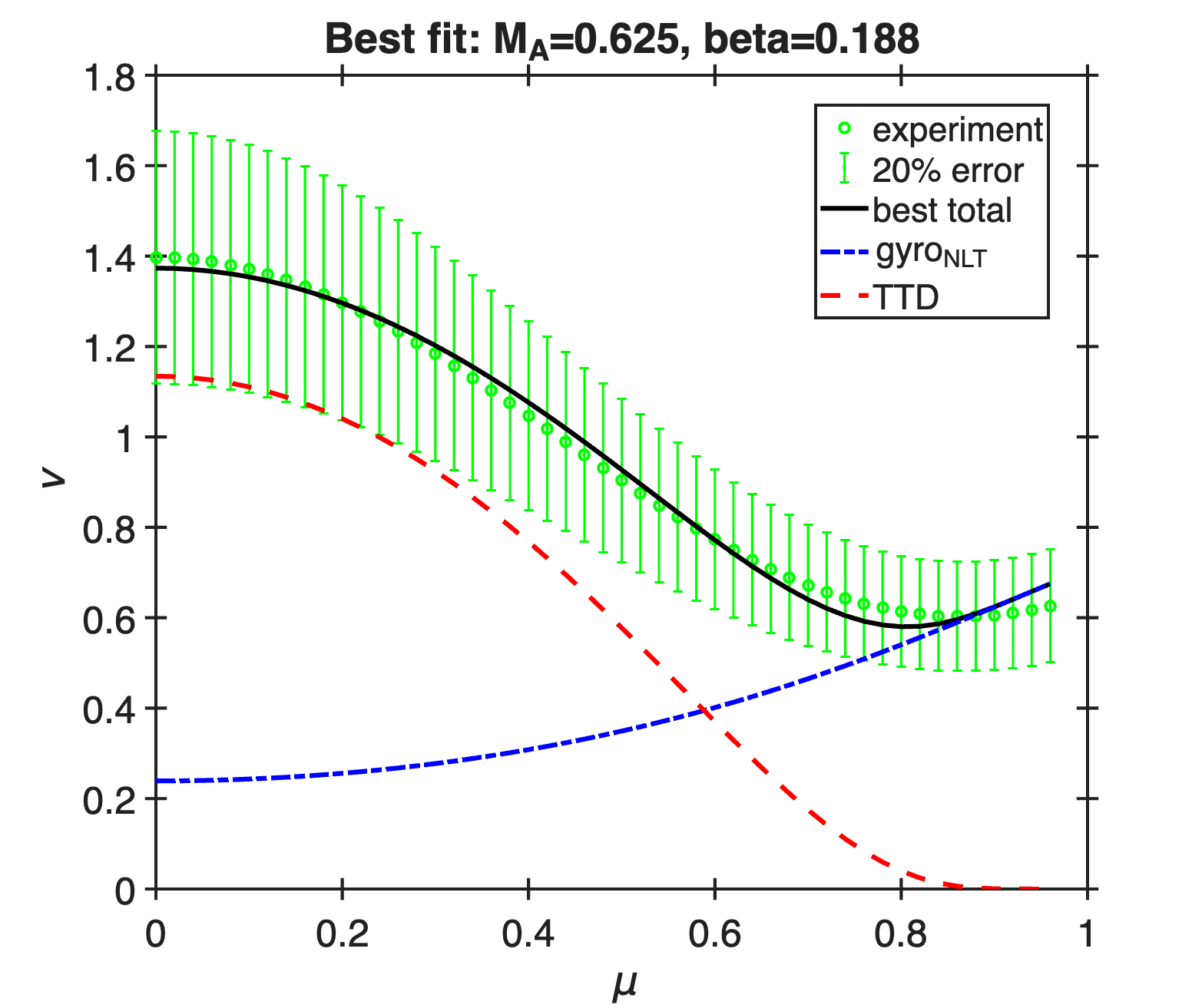}{0.31\textheight}{(b)}}
    \caption{(a) Contour map of RMS fractional deviation of the fitting between the theoretical and experimental curves. The black iso-contours mark the values from $5\%$ to $20\%$. The $\star$ marks the point with the minimum RMS deviation. (b) The best fitting with the corresponding parameters.}
    \label{fig:scan_results}
\end{figure*}
}
The agreement between theoretical results and observationally inferred values is achieved without any particular fine-tuning, lending direct support to the prediction of fast modes dominating CR scattering formulated in the NLT theory framework and placing a reasonable constraint on the environmental conditions of the LISM magnetic field and turbulence.

The dependence of the scattering rate on the pitch-angle cosine $\mu$ at higher particle energies is expected to steepen, owing to the more rapid decrease of the gyroresonance efficiency with increase of energy. The diffusion coefficient is weakly dependent on the energy/rigidity of CRs in the TeV energy range. Because of the slow drop of $D_{\mu\mu}$ with $\mu$ (Fig.~\ref{fig:comparison}a), the diffusion coefficient calculated by
\begin{equation}
    \lambda=\int \frac{v(1-\mu^2)^2}{D_{\mu\mu}} d\mu
\end{equation}
is dominated by $D_{\mu\mu}$ at large pitch angles through the mirror interaction, which is independent of energy, as shown in Eqs.~(\ref{eq:TTD_mu}, \ref{eq:TTD_ana}). However, as energy increases, gyroresonance falls more quickly, thereby enhancing their relative contribution at small pitch angles down to $\lambda$. Consequently, the rigidity dependence of the diffusion coefficient becomes progressively steeper at higher energies. 

\section{Discussion}

The 3D structure of turbulence is crucial for understanding the pattern of pitch angle scattering, particularly the dependence on $\mu$/pitch angle. The dependence of the TeV CR pitch-angle distribution is incompatible with traditional isotropic turbulence or slab waves. We find that compressible turbulence with a quasi-slab structure, which naturally arises from fast modes in low $\beta$ environments, provides the best explanation for the observed pitch-angle diffusion behavior. Firstly, the current understanding of the physical parameters in LISM is consistent with the low $\beta$ condition \citep{Zirnstein2016, ZhaoL2025}. Secondly, both mirror and gyroresonance should operate in order to reach the normal diffusion regime with a finite mean free path, which cannot be realized in either isotropic turbulence or slab waves. The slab modes can only give rise to gyroresonance, but not to the mirror/TTD interaction. Besides, the emergence of slab modes is likewise a challenge since self-confinement does not operate for CRs $>$ TeV due to both nonlinear Landau damping and turbulence damping \citep{Yan2002, FG04, Kulsrudbook, Lazarian2016}. The isotropic turbulence can, in principle, {\bf mirror/TTD scatter the cosmic rays if it is compressible modes}. The assumption of isotropy is not valid, though, due to anisotropic damping. Furthermore, the observed weak dependence on the pitch angle means that the mirroring effect must be reduced, disfavoring MHD modes with high obliquity. In other words, the turbulence must be quasi-parallel. 

One may argue that mirror interaction and gyroresonance can arise from different modes/structures in turbulence rather than from a single type of quasi-slab fast-mode turbulence. For instance, mirror/TTD interaction originates from slow modes or compressible modes truncated at scales much larger than the particle resonant scales owing to ion–neutral damping in a partially ionized plasma \cite{Yan2008,YLD2004}. Meanwhile, gyroresonant scattering may also be driven by a slab component generated through a separate process.\footnote{As discussed above, the generation of such a slab component is not straightforward for TeV cosmic rays.} In this scenario, extensive fine-tuning would then be inevitable in order to reproduce the observed smooth $\mu$ dependence of the pitch-angle scattering.

We also want to emphasize the difference between global CR transport and that occurring in the LISM. The pitch-angle diffusion coefficient derived from TeV CR anisotropies reflects mainly local propagation properties {since CRs quickly lose their memories of initial directions due to frequent scattering \citep[see][]{Giacinti2012, Kuhlen2022}}. There have been attempts to test the scattering coefficient obtained through microscopic physical processes directly against global CR observables, e.g., \citep{Kempski2022}. We argue that global and local transport can be completely different. For example, CR diffusion perpendicular to the local magnetic field is found to be small compared to parallel diffusion \citep{MZ2025}, whereas global CR transport indicates nearly isotropic diffusion. Similarly, reduced diffusion in local environments has been identified in other observations \citep{HAWC2017, LYZ19}. These are not necessarily contradictory, as the magnetic field structure at global scales differs substantially from that at local scales. On the global scale, the random walk of magnetic field lines can also contribute to CR spatial diffusion. 

Some recent work has advocated for potentially large-angle scattering by intermittent structure in MHD turbulence \citep{Lemoine_2023}. We do not consider it here for two reasons. First, there is no indication of superAlfvénic turbulence in LISM. The analysis of Voyager 1 data indicates the turbulence in the local bubble is trans-Alfvénic \citep{ZhaoL2025}. Secondly, there is no indication of large-angle scattering from either test-particle simulations in trans- or sub-Alfv\'enic turbulence or CR anisotropy observations \citep{YanK2026, MZ2025}. 

We recognize that the experimental result adopted from \cite{MZ2025} obtained by back-tracing, is not exclusive and is contingent upon the heliospheric magnetic structure selected for the modeling. Nevertheless, the concordance between the two entirely independent studies originating from the nonlinear theory on cosmic ray scattering by fast modes  \citep{Yan2008} and the back-tracing analysis in \cite{MZ2025}, spanning several decades, elucidates certain fundamental characteristics of the local interstellar turbulence and its interaction with cosmic rays.

\section{Conclusion}

We have revisited the calculation of CR scattering in LISM turbulence {\bf in the diffusion regime} based on the earlier NLT model from YL08 for compressible MHD turbulence, motivated by recent advances in both the understanding of interstellar turbulence and observations of cosmic-ray properties. Our main findings are as follows:

\begin{itemize}
    \item Gyroresonance and mirror/TTD interactions with fast-mode turbulence naturally provide pitch-angle scattering rates consistent with those inferred from TeV CR anisotropy measurements.
    \item The quasi-slab structure of fast modes resulting from damping in a low-$\beta$ medium is found to be responsible for a weak pitch-angle dependence of the scattering rate.
    \item The turbulence in the LISM should be compressible with an amplitude approximately $0.3\lesssim  \delta B/B_0 <1$. There is a certain degeneracy between the inferred $ \delta B/B_0$ and $\beta$ values. The plasma $\beta$ would be $\lesssim 0.2$ if the the amplitude of compressible fast modes is $\delta B/B_0 < 0.5$, consistent with the known plasma condition in the LISM.
\end{itemize}
The agreement between the theoretical and observational results confirms the NLT theory on the fast modes' dominance on CR scattering in turbulence, an essential piece of CR transport, and has broad implications for all related processes in high-energy astrophysics and the interstellar medium.

\begin{acknowledgments}
We are grateful to B.Q. Qiao and T.Z. Liu for their valuable discussions. S.Z. acknowledges the support from National Natural Science Foundation of China (NSFC) Excellent Young Scientists Fund (Overseas). The use of ChatGPT is acknowledged for assistance in refining the English grammar and sentence structure.
\end{acknowledgments}

\bibliography{refs}

@PREAMBLE{
 "\providecommand{\noopsort}[1]{}" 
 # "\providecommand{\singleletter}[1]{#1}%" 
}

@ARTICLE{2010ApJ...718L.194A,
	author = {{Abbasi}, R. and {Abdou}, Y. and {Abu-Zayyad}, T. and {Adams}, J. and 
	{Aguilar}, J.~A. and {Ahlers}, M. and {Andeen}, K. and {Auffenberg}, J. and 
	{Bai}, X. and {Baker}, M. and et al.},
	title = "{Measurement of the Anisotropy of Cosmic-ray Arrival Directions with IceCube}",
	journal = {apj},
	archivePrefix = "arXiv",
	eprint = {1005.2960},
	primaryClass = "astro-ph.HE",
	year = 2010,
	month = aug,
	volume = 718,
	pages = {L194-L198},
	doi = {10.1088/2041-8205/718/2/L194},
	adsurl = {http://adsabs.harvard.edu/abs/2010ApJ...718L.194A}
}

@ARTICLE{2019ApJ...871...96A,
       author = {{Abeysekara}, A.~U. and {Alfaro}, R. and {Alvarez}, C. and {Arceo}, R. and {Arteaga-Vel{\'a}zquez}, J.~C. and {Avila Rojas}, D. and {Belmont-Moreno}, E. and {BenZvi}, S.~Y. and {Brisbois}, C. and {Capistr{\'a}n}, T. and {Carramiana}, A. and {Casanova}, S. and {Cotti}, U. and {Cotzomi}, J. and {D{\'\i}az-V{\'e}lez}, J.~C. and {De Le{\'o}n}, C. and {De la Fuente}, E. and {Dichiara}, S. and {DuVernois}, M.~A. and {Espinoza}, C. and {Fiorino}, D.~W. and {Fleischhack}, H. and {Fraija}, N. and {Galv{\'a}n-G{\'a}mez}, A. and {Garc{\'\i}a-Gonz{\'a}lez}, J.~A. and {Gonz{\'a}lez}, M.~M. and {Goodman}, J.~A. and {Hampel-Arias}, Z. and {Harding}, J.~P. and {Hernandez}, S. and {Hona}, B. and {Hueyotl-Zahuantitla}, F. and {Iriarte}, A. and {Jardin-Blicq}, A. and {Joshi}, V. and {Lara}, A. and {Le{\'o}n Vargas}, H. and {Luis-Raya}, G. and {Malone}, K. and {Marinelli}, S.~S. and {Mart{\'\i}nez-Castro}, J. and {Martinez}, O. and {Matthews}, J.~A. and {Miranda-Romagnoli}, P. and {Moreno}, E. and {Mostaf{\'a}}, M. and {Nellen}, L. and {Newbold}, M. and {Nisa}, M.~U. and {Noriega-Papaqui}, R. and {P{\'e}rez-P{\'e}rez}, E.~G. and {Pretz}, J. and {Ren}, Z. and {Rho}, C.~D. and {Rivi{\`e}re}, C. and {Rosa-Gonz{\'a}lez}, D. and {Rosenberg}, M. and {Salazar}, H. and {Salesa Greus}, F. and {Sandoval}, A. and {Schneider}, M. and {Schoorlemmer}, H. and {Sinnis}, G. and {Smith}, A.~J. and {Surajbali}, P. and {Taboada}, I. and {Tollefson}, K. and {Torres}, I. and {Villaseor}, L. and {Weisgarber}, T. and {Wood}, J. and {Zepeda}, A. and {Zhou}, H. and {{\'A}lvarez}, J.~D. and {HAWC Collaboration} and {Aartsen}, M.~G. and {Ackermann}, M. and {Adams}, J. and {Aguilar}, J.~A. and {Ahlers}, M. and {Ahrens}, M. and {Altmann}, D. and {Andeen}, K. and {Anderson}, T. and {Ansseau}, I. and {Anton}, G. and {Arg{\"u}elles}, C. and {Auffenberg}, J. and {Axani}, S. and {Backes}, P. and {Bagherpour}, H. and {Bai}, X. and {Barbano}, A. and {Barron}, J.~P. and {Barwick}, S.~W. and {Baum}, V. and {Bay}, R. and {Beatty}, J.~J. and {Becker Tjus}, J. and {Becker}, K. -H. and {BenZvi}, S. and {Berley}, D. and {Bernardini}, E. and {Besson}, D.~Z. and {Binder}, G. and {Bindig}, D. and {Blaufuss}, E. and {Blot}, S. and {Bohm}, C. and {B{\"o}rner}, M. and {Bos}, F. and {B{\"o}ser}, S. and {Botner}, O. and {Bourbeau}, E. and {Bourbeau}, J. and {Bradascio}, F. and {Braun}, J. and {Bretz}, H. -P. and {Bron}, S. and {Brostean-Kaiser}, J. and {Burgman}, A. and {Busse}, R.~S. and {Carver}, T. and {Cheung}, E. and {Chirkin}, D. and {Clark}, K. and {Classen}, L. and {Collin}, G.~H. and {Conrad}, J.~M. and {Coppin}, P. and {Correa}, P. and {Cowen}, D.~F. and {Cross}, R. and {Dave}, P. and {Day}, M. and {de Andr{\'e}}, J.~P.~A.~M. and {De Clercq}, C. and {DeLaunay}, J.~J. and {Dembinski}, H. and {Deoskar}, K. and {De Ridder}, S. and {Desiati}, P. and {de Vries}, K.~D. and {de Wasseige}, G. and {de With}, M. and {DeYoung}, T. and {D{\'\i}az-V{\'e}lez}, J.~C. and {Dujmovic}, H. and {Dunkman}, M. and {Dvorak}, E. and {Eberhardt}, B. and {Ehrhardt}, T. and {Eichmann}, B. and {Eller}, P. and {Evenson}, P.~A. and {Fahey}, S. and {Fazely}, A.~R. and {Felde}, J. and {Filimonov}, K. and {Finley}, C. and {Franckowiak}, A. and {Friedman}, E. and {Fritz}, A. and {Gaisser}, T.~K. and {Gallagher}, J. and {Ganster}, E. and {Garrappa}, S. and {Gerhardt}, L. and {Ghorbani}, K. and {Giang}, W. and {Glauch}, T. and {Gl{\"u}senkamp}, T. and {Goldschmidt}, A. and {Gonzalez}, J.~G. and {Grant}, D. and {Griffith}, Z. and {Haack}, C. and {Hallgren}, A. and {Halve}, L. and {Halzen}, F. and {Hanson}, K. and {Hebecker}, D. and {Heereman}, D. and {Helbing}, K. and {Hellauer}, R. and {Hickford}, S. and {Hignight}, J. and {Hill}, G.~C. and {Hoffman}, K.~D. and {Hoffmann}, R. and {Hoinka}, T. and {Hokanson-Fasig}, B. and {Hoshina}, K. and {Huang}, F. and {Huber}, M. and {Hultqvist}, K. and {H{\"u}nnefeld}, M. and {Hussain}, R. and {In}, S. and {Iovine}, N. and {Ishihara}, A. and {Jacobi}, E. and {Japaridze}, G.~S. and {Jeong}, M. and {Jero}, K. and {Jones}, B.~J.~P. and {Kalaczynski}, P. and {Kang}, W. and {Kappes}, A. and {Kappesser}, D. and {Karg}, T. and {Karle}, A. and {Katz}, U. and {Kauer}, M. and {Keivani}, A. and {Kelley}, J.~L. and {Kheirandish}, A. and {Kim}, J. and {Kintscher}, T. and {Kiryluk}, J. and {Kittler}, T. and {Klein}, S.~R. and {Koirala}, R. and {Kolanoski}, H. and {K{\"o}pke}, L. and {Kopper}, C. and {Kopper}, S. and {Koskinen}, D.~J. and {Kowalski}, M. and {Krings}, K. and {Kroll}, M. and {Kr{\"u}ckl}, G. and {Kunwar}, S. and {Kurahashi}, N. and {Kyriacou}, A. and {Labare}, M. and {Lanfranchi}, J.~L. and {Larson}, M.~J. and {Lauber}, F. and {Leonard}, K. and {Leuermann}, M. and {Liu}, Q.~R. and {Lohfink}, E. and {Lozano Mariscal}, C.~J. and {Lu}, L. and {L{\"u}nemann}, J. and {Luszczak}, W. and {Madsen}, J. and {Maggi}, G. and {Mahn}, K.~B.~M. and {Makino}, Y. and {Mancina}, S. and {Mari{\c{s}}}, I.~C. and {Maruyama}, R. and {Mase}, K. and {Maunu}, R. and {Meagher}, K. and {Medici}, M. and {Meier}, M. and {Menne}, T. and {Merino}, G. and {Meures}, T. and {Miarecki}, S. and {Micallef}, J. and {Moment{\'e}}, G. and {Montaruli}, T. and {Moore}, R.~W. and {Moulai}, M. and {Nagai}, R. and {Nahnhauer}, R. and {Nakarmi}, P. and {Naumann}, U. and {Neer}, G. and {Niederhausen}, H. and {Nowicki}, S.~C. and {Nygren}, D.~R. and {Obertacke Pollmann}, A. and {Olivas}, A. and {O'Murchadha}, A. and {O'Sullivan}, E. and {Palczewski}, T. and {Pandya}, H. and {Pankova}, D.~V. and {Peiffer}, P. and {Pepper}, J.~A. and {P{\'e}rez de los Heros}, C. and {Pieloth}, D. and {Pinat}, E. and {Pizzuto}, A. and {Plum}, M. and {Price}, P.~B. and {Przybylski}, G.~T. and {Raab}, C. and {Rameez}, M. and {Rauch}, L. and {Rawlins}, K. and {Rea}, I.~C. and {Reimann}, R. and {Relethford}, B. and {Renzi}, G. and {Resconi}, E. and {Rhode}, W. and {Richman}, M. and {Robertson}, S. and {Rongen}, M. and {Rott}, C. and {Ruhe}, T. and {Ryckbosch}, D. and {Rysewyk}, D. and {Safa}, I. and {Sanchez Herrera}, S.~E. and {Sandrock}, A. and {Sandroos}, J. and {Santander}, M. and {Sarkar}, S. and {Sarkar}, S. and {Satalecka}, K. and {Schaufel}, M. and {Schlunder}, P. and {Schmidt}, T. and {Schneider}, A. and {Schneider}, J. and {Sch{\"o}neberg}, S. and {Schumacher}, L. and {Sclafani}, S. and {Seckel}, D. and {Seunarine}, S. and {Soedingrekso}, J. and {Soldin}, D. and {Song}, M. and {Spiczak}, G.~M. and {Spiering}, C. and {Stachurska}, J. and {Stamatikos}, M. and {Stanev}, T. and {Stasik}, A. and {Stein}, R. and {Stettner}, J. and {Steuer}, A. and {Stezelberger}, T. and {Stokstad}, R.~G. and {St{\"o}{\ss}l}, A. and {Strotjohann}, N.~L. and {Stuttard}, T. and {Sullivan}, G.~W. and {Sutherland}, M. and {Taboada}, I. and {Tenholt}, F. and {Ter-Antonyan}, S. and {Terliuk}, A. and {Tilav}, S. and {Toale}, P.~A. and {Tobin}, M.~N. and {T{\"o}nnis}, C. and {Toscano}, S. and {Tosi}, D. and {Tselengidou}, M. and {Tung}, C.~F. and {Turcati}, A. and {Turcotte}, R. and {Turley}, C.~F. and {Ty}, B. and {Unger}, E. and {Unland Elorrieta}, M.~A. and {Usner}, M. and {Vandenbroucke}, J. and {Van Driessche}, W. and {van Eijk}, D. and {van Eijndhoven}, N. and {Vanheule}, S. and {van Santen}, J. and {Vraeghe}, M. and {Walck}, C. and {Wallace}, A. and {Wallraff}, M. and {Wandler}, F.~D. and {Wandkowsky}, N. and {Watson}, T.~B. and {Weaver}, C. and {Weiss}, M.~J. and {Wendt}, C. and {Werthebach}, J. and {Westerhoff}, S. and {Whelan}, B.~J. and {Whitehorn}, N. and {Wiebe}, K. and {Wiebusch}, C.~H. and {Wille}, L. and {Williams}, D.~R. and {Wills}, L. and {Wolf}, M. and {Wood}, J. and {Wood}, T.~R. and {Woolsey}, E. and {Woschnagg}, K. and {Wrede}, G. and {Xu}, D.~L. and {Xu}, X.~W. and {Xu}, Y. and {Yanez}, J.~P. and {Yodh}, G. and {Yoshida}, S. and {Yuan}, T. and {IceCube Collaboration}},
        title = "{All-sky Measurement of the Anisotropy of Cosmic Rays at 10 TeV and Mapping of the Local Interstellar Magnetic Field}",
      journal = {\apj},
         year = 2019,
        month = jan,
       volume = {871},
       number = {1},
          eid = {96},
        pages = {96},
          doi = {10.3847/1538-4357/aaf5cc},
archivePrefix = {arXiv},
       eprint = {1812.05682},
 primaryClass = {astro-ph.HE},
       adsurl = {https://ui.adsabs.harvard.edu/abs/2019ApJ...871...96A}
}

@ARTICLE{2010ApJ...711..119A,
       author = {{Amenomori}, M. and {Bi}, X.~J. and {Chen}, D. and {Cui}, S.~W. and {Danzengluobu} and {Ding}, L.~K. and {Ding}, X.~H. and {Fan}, C. and {Feng}, C.~F. and {Feng}, Zhaoyang and {Feng}, Z.~Y. and {Gao}, X.~Y. and {Geng}, Q.~X. and {Gou}, Q.~B. and {Guo}, H.~W. and {He}, H.~H. and {He}, M. and {Hibino}, K. and {Hotta}, N. and {Hu}, Haibing and {Hu}, H.~B. and {Huang}, J. and {Huang}, Q. and {Jia}, H.~Y. and {Jiang}, L. and {Kajino}, F. and {Kasahara}, K. and {Katayose}, Y. and {Kato}, C. and {Kawata}, K. and {Labaciren} and {Le}, G.~M. and {Li}, A.~F. and {Li}, H.~C. and {Li}, J.~Y. and {Liu}, C. and {Lou}, Y. -Q. and {Lu}, H. and {Meng}, X.~R. and {Mizutani}, K. and {Mu}, J. and {Munakata}, K. and {Nagai}, A. and {Nanjo}, H. and {Nishizawa}, M. and {Ohnishi}, M. and {Ohta}, I. and {Ozawa}, S. and {Saito}, T. and {Saito}, T.~Y. and {Sakata}, M. and {Sako}, T.~K. and {Shibata}, M. and {Shiomi}, A. and {Shirai}, T. and {Sugimoto}, H. and {Takita}, M. and {Tan}, Y.~H. and {Tateyama}, N. and {Torii}, S. and {Tsuchiya}, H. and {Udo}, S. and {Wang}, B. and {Wang}, H. and {Wang}, Y. and {Wang}, Y.~G. and {Wu}, H.~R. and {Xue}, L. and {Yamamoto}, Y. and {Yan}, C.~T. and {Yang}, X.~C. and {Yasue}, S. and {Ye}, Z.~H. and {Yu}, G.~C. and {Yuan}, A.~F. and {Yuda}, T. and {Zhang}, H.~M. and {Zhang}, J.~L. and {Zhang}, N.~J. and {Zhang}, X.~Y. and {Zhang}, Y. and {Zhang}, Yi and {Zhang}, Ying and {Zhaxisangzhu} and {Zhou}, X.~X. and {Tibet AS{\ensuremath{\gamma}} Collaboration}},
        title = "{On Temporal Variations of the Multi-TeV Cosmic Ray Anisotropy Using the Tibet III Air Shower Array}",
      journal = {\apj},
         year = 2010,
        month = mar,
       volume = {711},
       number = {1},
        pages = {119-124},
          doi = {10.1088/0004-637X/711/1/119},
archivePrefix = {arXiv},
       eprint = {1001.2646},
 primaryClass = {astro-ph.HE},
       adsurl = {https://ui.adsabs.harvard.edu/abs/2010ApJ...711..119A}
}

@Article{Blasi2012,
	Title                    = {{Diffusive propagation of cosmic rays from supernova remnants in the Galaxy. II: anisotropy}},
	Author                   = {{Blasi}, P. and {Amato}, E.},
	Journal                  = {jcap},
	Year                     = {2012},
	
	Month                    = jan,
	Pages                    = {11},
	Volume                   = {1},
	Adsurl                   = {http://adsabs.harvard.edu/abs/2012JCAP...01..011B},
	Archiveprefix            = {arXiv},
	Doi                      = {10.1088/1475-7516/2012/01/011},
	Eid                      = {011},
	Eprint                   = {1105.4529},
	Primaryclass             = {astro-ph.HE}
}

@ARTICLE{Liu2017,
	author = {{Liu}, Wei and {Bi}, Xiao-Jun and {Lin}, Su-Jie and {Wang}, Bing-Bing and {Yin}, Peng-Fei},
	title = "{Excesses of cosmic ray spectra from a single nearby source}",
	journal = {prd},
	year = 2017,
	month = jul,
	volume = {96},
	number = {2},
	eid = {023006},
	pages = {023006},
	doi = {10.1103/PhysRevD.96.023006},
	archivePrefix = {arXiv},
	eprint = {1611.09118},
	primaryClass = {astro-ph.HE},
	adsurl = {https://ui.adsabs.harvard.edu/abs/2017PhRvD..96b3006L}
}

@ARTICLE{2017ApJ...842...54L,
       author = {{L{\'o}pez-Barquero}, V. and {Xu}, S. and {Desiati}, P. and {Lazarian}, A. and {Pogorelov}, N.~V. and {Yan}, H.},
        title = "{TeV Cosmic-Ray Anisotropy from the Magnetic Field at the Heliospheric Boundary}",
      journal = {\apj},
         year = 2017,
        month = jun,
       volume = {842},
       number = {1},
          eid = {54},
        pages = {54},
          doi = {10.3847/1538-4357/aa74d1},
archivePrefix = {arXiv},
       eprint = {1610.03097},
 primaryClass = {astro-ph.HE},
       adsurl = {https://ui.adsabs.harvard.edu/abs/2017ApJ...842...54L}
}

@article{Beresnyak2011,
author = {Beresnyak, A. and Yan, H. and Lazarian, A.},
doi = {10.1088/0004-637X/728/1/60},
journal = {Astrophys. J.},
pages = {60},
title = {{Numerical Study of Cosmic Ray Diffusion in Magnetohydrodynamic Turbulence}},
volume = {728},
year = {2011}
}

@article{Giacinti2012,
  title = {Local Magnetic Turbulence and TeV--PeV Cosmic Ray Anisotropies},
  author = {Giacinti, Gwenael and Sigl, G\"unter},
  journal = {Phys. Rev. Lett.},
  volume = {109},
  issue = {7},
  pages = {071101},
  numpages = {5},
  year = {2012},
  month = {Aug},
  publisher = {American Physical Society},
  doi = {10.1103/PhysRevLett.109.071101},
  url = {https://link.aps.org/doi/10.1103/PhysRevLett.109.071101}
}

@book{longair1997,
  title={High Energy Astrophysics: Volume 2, Stars, the Galaxy and the Interstellar Medium},
  author={Longair, Malcolm S},
  year={1997},
  publisher={Cambridge University Press},
  address={Cambridge}
}

@article{Qiao2026,
author = {Qiao, Bing-Qiang and Liu, Wei and Yan, Huirong and Guo, Yi-Qing},
doi = {10.3847/1538-4357/ae2015},
journal = {Astrophys. J.},
pages = {77},
title = {{The Compton–Getting Origin of the Large-scale Anisotropy of Galactic Cosmic Rays}},
volume = {996},
year = {2026}
}

@article{Kuhlen2022,
doi = {10.3847/1538-4357/ac503b},
url = {https://doi.org/10.3847/1538-4357/ac503b},
year = {2022},
month = {mar},
publisher = {The American Astronomical Society},
volume = {927},
number = {1},
pages = {110},
author = {Kuhlen, Marco and Phan, Vo Hong Minh and Mertsch, Philipp},
title = {No Longer Ballistic, Not Yet Diffusive‚Äîthe Formation of Cosmic-Ray Small-scale Anisotropies},
journal = {The Astrophysical Journal}
}

@article{Felice2001,
author = {Felice, G. M. and Kulsrud, R. M.},
doi = {10.1086/320651},
journal = {Astrophys. J.},
pages = {198--210},
title = {{Cosmic‐Ray Pitch‐Angle Scattering through 90o}},
volume = {553},
year = {2001}
}

@article{Yan2008,
author = {Yan, Huirong and Lazarian, A.},
title = {{Cosmic‐Ray Propagation: Nonlinear Diffusion Parallel and Perpendicular to Mean Magnetic Field}},
journal = {Astrophys. J.},
doi = {10.1086/524771},
pages = {942--953},
volume = {673},
year = {2008}
}

@article{GS95,
author = {Goldreich, P. and Sridhar, S.},
title = {{Toward a theory of interstellar turbulence. 2: Strong alfvenic turbulence}},
journal = {Astrophys. J.},
pages = {763},
volume = {438},
year = {1995},
doi = {10.1086/175121}
}

@article{Makwana2020,
author = {Makwana, K. D. and Yan, Huirong},
title = {{Properties of Magnetohydrodynamic Modes in Compressively Driven Plasma Turbulence}},
journal = {Phys. Rev. X},
pages = {031021},
volume = {10},
year = {2020},
doi = {10.1103/PhysRevX.10.031021}
}

@article{CL02,
author = {Cho, Jungyeon and Lazarian, A.},
title = {{Compressible Sub-Alfv{\'{e}}nic MHD Turbulence in Low-$\beta$ Plasmas}},
journal = {Phys. Rev. Lett.},
pages = {245001},
volume = {88},
year = {2002},
doi = {10.1103/PhysRevLett.88.245001}
}

@article{CL03,
author = {Cho, Jungyeon and Lazarian, A.},
title = {{Compressible magnetohydrodynamic turbulence: Mode coupling, scaling relations, anisotropy, viscosity-damped regime and astrophysical implications}},
journal = {MNRAS},
pages = {325--339},
volume = {345},
year = {2003},
doi = {10.1098/rsta.2014.0152}
}

@article{Suzuki2006,
author = {Suzuki, T. K. and Yan, H. and Lazarian, A. and et al.},
title = {{Collisionless Damping of Fast Magnetohydrodynamic Waves in Magnetorotational Winds}},
journal = {Astrophys. J.},
pages = {1005--1017},
volume = {640},
year = {2006},
doi = {10.1086/500164}
}

@article{Hadid2018,
author = {Hadid, L. Z. and Sahraoui, F. and Galtier, S. and et al.},
title = {{Compressible Magnetohydrodynamic Turbulence in the Earth's Magnetosheath: Estimation of the Energy Cascade Rate Using in situ Spacecraft Data}},
journal = {Phys. Rev. Lett.},
pages = {055102},
volume = {120},
year = {2018}
}

@article{Yan2004,
author = {Yan, Huirong and Lazarian, A.},
title = {{Cosmic‐Ray Scattering and Streaming in Compressible Magnetohydrodynamic Turbulence}},
journal = {Astrophys. J.},
pages = {757--769},
volume = {614},
year = {2004},
doi = {10.1086/423733},
}

@article{Petrosian2006,
author = {Petrosian, Vahe and Yan, Huirong and Lazarian, A},
title = {{Damping of Magnetohydrodynamic Turbulence in Solar Flares}},
journal = {Astrophys. J.},
pages = {603--612},
volume = {644},
year = {2006},
doi = {10.1086/503378}
}

@article{Zhao2024ApJ,
author = {Zhao, Siqi and Yan, Huirong and Liu, Terry Z. and Yuen, Ka Ho and Shi, Mijie},
doi = {10.3847/1538-4357/ad132e},
journal = {Astrophys. J.},
pages = {89},
title = {{Small-amplitude Compressible Magnetohydrodynamic Turbulence Modulated by Collisionless Damping in Earth's Magnetosheath: Observation Matches Theory}},
volume = {962},
year = {2024}
}

@INPROCEEDINGS{Yan2022,
       author = {{Yan}, H.},
    booktitle = {37th International Cosmic Ray Conference},
         year = 2022,
        month = mar,
          eid = {38},
        pages = {38},
          doi = {10.22323/1.395.0038},
archivePrefix = {arXiv},
       eprint = {2109.07847},
 primaryClass = {astro-ph.HE},
       adsurl = {https://ui.adsabs.harvard.edu/abs/2022icrc.confE..38Y}
}

@article{Iroshnikov,
author = {Iroshnikov, P},
title = {{Turbulence of a Conducting Fluid in a Strong Magnetic Field}},
journal = {Astron. Zh.},
pages = {742},
volume = {40},
year = {1963}
}

@article{Galtier23, volume={89}, journal={J. Plasma Phys.},author={Galtier, Sébastien}, title={Fast magneto-acoustic wave turbulence and the Iroshnikov–Kraichnan spectrum}, year={2023}, pages={905890205}}

@article{Yan2002,
author = {Yan, Huirong and Lazarian, A.},
doi = {10.1103/PhysRevLett.89.281102},
journal = {Phys. Rev. Lett.},
pages = {281102},
title = {{Scattering of Cosmic Rays by Magnetohydrodynamic Interstellar Turbulence}},
volume = {89},
year = {2002}
}

@article{Maiti2022,
author = {Maiti, Snehanshu and Makwana, Kirit and Zhang, Heshou and Yan, Huirong},
doi = {10.3847/1538-4357/ac46c8},
journal = {Astrophys. J.},
pages = {94},
publisher = {IOP Publishing},
title = {{Cosmic-ray Transport in Magnetohydrodynamic Turbulence}},
volume = {926},
year = {2022}
}

@article{Jokipii1966,
author = {Jokipii, J. R.},
journal = {Astrophys. J.},
pages = {480},
title = {{Cosmic-Ray Propagation. I. Charged Particles in a Random Magnetic Field}},
volume = {146},
year = {1966},
doi = {10.1086/148912}
}

@book{Schlickeiser2002,
author = {Schlickeiser, Reinhard},
booktitle = {Theoretical Physics and Astrophysics},
doi = {10.1007/978-3-662-04814-6},
pages = {343--387},
publisher = {Springer Berlin Heidelberg},
series = {Astronomy and Astrophysics Library},
title = {{Cosmic Ray Astrophysics}},
year = {2002}
}

@article{FG04,
author = {Farmer, Alison J. and Goldreich, Peter},
doi = {10.1086/382040},
journal = {Astrophys. J.},
pages = {671--674},
title = {{Wave Damping by Magnetohydrodynamic Turbulence and Its Effect on Cosmic‐Ray Propagation in the Interstellar Medium}},
volume = {604},
year = {2004}
}

@article{Schlickeiser1993,
author={Schlickeiser, R. and Achatz, U.},
title={Cosmic-ray particle transport in weakly turbulent plasmas. Part 1. Theory}, 
volume={49}, 
doi={10.1017/S0022377800016822},
journal={Journal of Plasma Physics},
year={1993},
pages={63–77}
}

@article{YLD2004,
doi = {10.1086/425111},
url = {https://doi.org/10.1086/425111},
year = {2004},
month = {dec},
publisher = {},
volume = {616},
number = {2},
pages = {895},
author = {Yan, Huirong and Lazarian, A. and Draine, B. T.},
title = {Dust Dynamics in Compressible Magnetohydrodynamic Turbulence},
journal = {The Astrophysical Journal}
}

@article{Yan2011,
author = {Yan, Huirong and Lazarian, A.},
doi = {10.1088/0004-637X/731/1/35},
journal = {Astrophys. J.},
pages = {35},
title = {{Cosmic Ray transport through gyroresonance instability in compressible turbulence}},
volume = {731},
year = {2011}
}

@misc{YanK2026,
      title={Non-Markovian Cosmic-Ray Pitch-Angle Transport from Mirror Interactions}, 
      author={Kai Yan and Huirong Yan and Parth Pavaskar and Chuanpeng Hou and Ruo-Yu Liu},
      year={2026},
      eprint={2603.19037},
      archivePrefix={arXiv},
      primaryClass={astro-ph.HE},
      url={https://arxiv.org/abs/2603.19037}, 
}

@article{Lebiga2018,
author = {Lebiga, O. and Santos-Lima, R. and Yan, H.},
doi = {10.1093/mnras/sty309},
journal = {MNRAS},
pages = {2779--2791},
title = {{Kinetic-MHD simulations of gyroresonance instability driven by CR pressure anisotropy}},
volume = {476},
year = {2018}
}

@article{Bykov2013,
author = {Bykov, A. M. and Brandenburg, A. and Malkov, M. A. and Osipov, S. M.},
doi = {10.1007/s11214-013-9988-3},
journal = {Space Science Reviews},
month = {oct},
number = {2-4},
pages = {201--232},
title = {{Microphysics of Cosmic Ray Driven Plasma Instabilities}},
volume = {178},
year = {2013}
}

@article{Chandran2000,
	adsurl = {http://adsabs.harvard.edu/abs/2000PhRvL..85.4656C},
	author = {{Chandran}, B.~D.~G.},
	doi = {10.1103/PhysRevLett.85.4656},
	journal = {Phys. Rev. Lett.},
	month = nov,
	pages = {4656-4659},
	title = {{Scattering of Energetic Particles by Anisotropic Magnetohydrodynamic Turbulence with a Goldreich-Sridhar Power Spectrum}},
	volume = 85,
	year = 2000}

@article{ginzburg1962,
  title={Propagation of electromagnetic waves in plasma},
  author={Ginzburg, Vitaliĭ, Lazarevich and Sadowski, Walter L and Gallik, DM and Brown, Sanborn C},
  journal={Phys. Today},
  volume={15},
  number={10},
  pages={70--73},
  year={1962},
  publisher={American Institute of Physics}
}

@article{Lee1973,
   author = {Martin A. Lee and Heinrich J. Völk},
   doi = {10.1007/BF00648673/METRICS},
   issn = {0004640X},
   issue = {1},
   journal = {Astrophysics and Space Science},
   month = {9},
   pages = {31-49},
   publisher = {Kluwer Academic Publishers},
   title = {Damping and nonlinear wave-particle interactions of Alfvén-waves in the solar wind},
   volume = {24},
   url = {https://link.springer.com/article/10.1007/BF00648673},
   year = {1973}
}

@article{ZhangM2020,
   author = {Ming Zhang and N. V. Pogorelov and Y. Zhang and H. B. Hu and R. Schlickeiser},
   doi = {10.3847/1538-4357/ab643c},
   issn = {0004-637X},
   issue = {2},
   journal = {Astrophys. J.},
   month = {2},
   pages = {97},
   publisher = {American Astronomical Society},
   title = {The Original Anisotropy of TeV Cosmic Rays in the Local Interstellar Medium},
   volume = {889},
   year = {2020}
}

@article{MZ2025,
   author = {N. D. Maalal and M. Zhang},
   doi = {10.3847/1538-4357/adfc52},
   issn = {0004-637X},
   issue = {1},
   journal = {Astrophys. J.},
   month = {10},
   pages = {46},
   title = {Pitch-angle Focusing and Scattering of TeV Galactic Cosmic Rays Propagating in Local Interstellar Magnetic Fields},
   volume = {992},
   url = {https://iopscience.iop.org/article/10.3847/1538-4357/adfc52},
   year = {2025}
}

@article{Hou2025,
author = {Hou, Chuanpeng and Yan, Huirong and Zhao, Siqi and Pavaskar, Parth},
doi = {10.3847/2041-8213/ae0c97},
journal = {The Astrophysical Journal Letters},
pages = {L28},
publisher = {IOP Publishing},
title = {{Energy Cascade and Damping in Fast-mode Compressible Turbulence}},
volume = {992},
year = {2025}
}

@article{Hou2026,
doi = {10.3847/1538-4357/ae6f0a},
url = {https://doi.org/10.3847/1538-4357/ae6f0a},
year = {2026},
month = {jun},
publisher = {The American Astronomical Society},
volume = {1004},
number = {1},
pages = {64},
author = {Hou, Chuanpeng and Yan, Huirong and Zhao, Siqi},
title = {Compressible Turbulence as a Source of Particle Beams and Ion Bernstein Waves in Collisionless Plasmas},
journal = {The Astrophysical Journal}
}

@article{Volk1975,
   author = {Heinrich J. Völk},
   doi = {10.1029/RG013i004p00547},
   issn = {19449208},
   issue = {4},
   journal = {Reviews of Geophysics},
   pages = {547-566},
   title = {Cosmic ray propagation in interplanetary space},
   volume = {13},
   year = {1975}
}

@article{Ahlers2017,
title = {Origin of small-scale anisotropies in Galactic cosmic rays},
journal = {Progress in Particle and Nuclear Physics},
volume = {94},
pages = {184-216},
year = {2017},
issn = {0146-6410},
doi = {https://doi.org/10.1016/j.ppnp.2017.01.004},
url = {https://www.sciencedirect.com/science/article/pii/S0146641017300054},
author = {Markus Ahlers and Philipp Mertsch}
}

@article{Zirnstein2016,
doi = {10.3847/2041-8205/818/1/L18},
url = {https://doi.org/10.3847/2041-8205/818/1/L18},
year = {2016},
month = {feb},
publisher = {The American Astronomical Society},
volume = {818},
number = {1},
pages = {L18},
author = {Zirnstein, E. J. and Heerikhuisen, J. and Funsten, H. O. and Livadiotis, G. and McComas, D. J. and Pogorelov, N. V.},
title = {LOCAL INTERSTELLAR MAGNETIC FIELD DETERMINED FROM THE INTERSTELLAR BOUNDARY EXPLORER RIBBON},
journal = {The Astrophysical Journal Letters}
}

@article{ZhaoL2025,
   author = {Zhao, L.-L. and Florinski, V. and Zank, G. P. and Opher, M. and Richardson, J. and Kurth, W. S. and Silwal, A. and Zhu, X. and Subashchandar, N. S. M. and Alonso Guzman, J. G. and Jin, Z.},
   doi = {10.3847/2041-8213/ae09aa},
   issn = {2041-8205},
   issue = {1},
   journal = {The Astrophysical Journal Letters},
   month = {10},
   pages = {L4},
   publisher = {IOP Publishing},
   title = {Magnetic Turbulence Intermittency and Compressibility in the Inner Heliosheath and Very Local Interstellar Medium},
   volume = {992},
   year = {2025}
}

@article{Lazarian2009,
author = {Lazarian, A and Beresnyak, A and Yan, H and Opher, M and Liu, Y},
doi = {10.1007/s11214-008-9452-y},
journal = {Space Science Reviews},
number = {1},
pages = {387--413},
title = {{Properties and Selected Implications of Magnetic Turbulence for Interstellar Medium, Local Bubble and Solar Wind}},
url = {https://doi.org/10.1007/s11214-008-9452-y},
volume = {143},
year = {2009}
}

@article{Kempski2022,
   author = {Philipp Kempski and Eliot Quataert},
   doi = {10.1093/mnras/stac1240},
   issn = {13652966},
   issue = {1},
   journal = {Monthly Notices of the Royal Astronomical Society},
   month = {6},
   pages = {657-674},
   publisher = {Oxford Academic},
   title = {Reconciling cosmic ray transport theory with phenomenological models motivated by Milky-Way data},
   volume = {514},
   url = {https://dx.doi.org/10.1093/mnras/stac1240},
   year = {2022}
}

@article{HAWC2017,
	adsurl = {http://adsabs.harvard.edu/abs/2017Sci...358..911A},
	archiveprefix = {arXiv},
	author = {{Abeysekara}, A.~U. and {Albert}, A. and {Alfaro}, R. and {Alvarez}, C. and {{\'A}lvarez}, J.~D. and et al.},
	doi = {10.1126/science.aan4880},
	journal = {Science},
	pages = {911-914},
	primaryclass = {astro-ph.HE},
	title = {{Extended gamma-ray sources around pulsars constrain the origin of the positron flux at Earth}},
	volume = {358},
	year = {2017}}

@book{Kulsrudbook,
	adsurl = {http://adsabs.harvard.edu/abs/2005ppfa.book.....K},
	author = {{Kulsrud}, R.~M.},
	booktitle = {Plasma physics for astrophysics / Russell M.~Kulsrud.~Princeton, N.J.~},
    publisher = {Princeton University Press, c2005.~(Princeton series in astrophysics)},
	title = {{Plasma physics for astrophysics}},
	year = 2005}

@article{LYZ19,
	author = {Liu, Ruo-Yu and Yan, Huirong and Zhang, Heshou},
	doi = {10.1103/PhysRevLett.123.221103},
	issue = {22},
	journal = {Phys. Rev. Lett.},
	month = {Nov},
	numpages = {5},
	pages = {221103},
	publisher = {American Physical Society},
	title = {Understanding the Multiwavelength Observation of Geminga's Tev Halo: The Role of Anisotropic Diffusion of Particles},
	url = {https://link.aps.org/doi/10.1103/PhysRevLett.123.221103},
	volume = {123},
	year = {2019}}

@article{Lazarian2016,
   author = {A. Lazarian},
   doi = {10.3847/1538-4357/833/2/131},
   issn = {0004-637X},
   issue = {2},
   journal = {Astrophys. J.},
   month = {12},
   pages = {131},
   publisher = {IOP Publishing},
   title = {DAMPING OF ALFV\'EN WAVES BY TURBULENCE AND ITS CONSEQUENCES: FROM COSMIC-RAY STREAMING TO LAUNCHING WINDS},
   volume = {833},
   url = {https://iopscience.iop.org/article/10.3847/1538-4357/833/2/131 https://iopscience.iop.org/article/10.3847/1538-4357/833/2/131/meta},
   year = {2016}
}

@article{Zhao2026,
author = {Zhao, Siqi and Yan, Huirong and Liu, Terry Z. and Hou, Chuanpeng and Yuen, Ka Ho},
doi = {10.3847/1538-4357/ae7e87},
journal = {The Astrophysical Journal},
pages = {182},
title = {{Spatiotemporal Properties of Compressible Magnetohydrodynamic Turbulence from Space Plasma}},
url = {https://iopscience.iop.org/article/10.3847/1538-4357/ae7e87},
volume = {1005},
year = {2026}
}

@article{Zhao2026_slab,
author = {Zhao, Siqi and Yan, Huirong and Liu, Terry Z. and Hou, Chuanpeng},
doi = {10.3847/1538-4357/ae2866},
journal = {Astrophys. J.},
pages = {46},
publisher = {IOP Publishing},
title = {{Mode Composition Shapes Magnetic Anisotropy in Solar Wind Turbulence}},
volume = {996},
year = {2026}
}

@article{Yuen2025,
author = {Yuen, Ka Ho and Li, Hui and Yan, Huirong},
doi = {10.3847/1538-4357/add3ee},
journal = {Astrophys. J.},
pages = {221},
publisher = {IOP Publishing},
title = {{Temporal Properties of Compressible Magnetohydrodynamic Turbulence}},
volume = {986},
year = {2025}
}

@article{Zhao2022,
author = {Zhao, S. Q. and Yan, Huirong and Liu, Terry Z. and Liu, Mingzhe and Wang, Huizi},
doi = {10.3847/1538-4357/ac822e},
journal = {Astrophys. J.},
number = {2},
pages = {102},
title = {{Multispacecraft Analysis of the Properties of Magnetohydrodynamic Fluctuations in Sub-Alfv{\'{e}}nic Solar Wind Turbulence at 1 au}},
volume = {937},
year = {2022}
}

@article{Lemoine_2023, 
        title={Particle transport through localized interactions with sharp magnetic field bends in MHD turbulence},
        volume={89}, 
        DOI={10.1017/S0022377823000946},
        journal={Journal of Plasma Physics}, author={Lemoine, Martin}, 
        year={2023}, pages={175890501}
        }

@article{Kraichnan1965,
author = {Kraichnan, Robert H},
doi = {10.1063/1.1761412},
journal = {The Physics of Fluids},
month = {jul},
number = {7},
pages = {1385--1387},
title = {{Inertial-Range Spectrum of Hydromagnetic Turbulence}},
volume = {8},
year = {1965}
}

@article{Chandran2005,
author = {Chandran, Benjamin D.G.},
doi = {10.1103/PhysRevLett.95.265004},
journal = {Physical Review Letters},
number = {26},
pages = {1--4},
title = {{Weak compressible magnetohydrodynamic turbulence in the solar corona}},
volume = {95},
year = {2005}
}
\end{document}